\documentstyle[12pt,aaspp4,tighten]{article}

\newcommand{\lsun}{log $L/L_{\odot}\,$}
\newcommand{\msun}{$M/M_{\odot}\,$}

\newcommand{\seco}{second overtone~}
\newcommand{\ds}{$\delta$ Scuti~}
\newcommand{\so}{SO~}

\begin{document}

\title{SECOND OVERTONE PULSATORS AMONG $\delta$ SCUTI STARS}

\author{Giuseppe Bono}
\affil{Osservatorio Astronomico di Trieste, Via G.B. Tiepolo 11,
34131 Trieste, Italy; bono@oat.ts.astro.it}

\author{Filippina Caputo}
\affil{Osservatorio Astronomico di Capodimonte, Via Moiariello 16,
80131 Napoli, Italy; caputo@astrna.na.astro.it}

\author{Santi Cassisi \altaffilmark{1}}
\affil{Osservatorio Astronomico di Teramo, Via M. Maggini,
64100 Teramo, Italy; cassisi@astrte.te.astro.it}

\author{Vittorio Castellani \altaffilmark{2}}
\affil{Dipartimento di Fisica, Universit\`a di Pisa, Piazza Torricelli 2,
56100 Pisa, Italy; vittorio@astr1pi.difi.unipi.it}

\author{Marcella Marconi}
\affil{Dipartimento di Fisica, Universit\`a di Pisa, Piazza Torricelli 2,
56100 Pisa, Italy; marcella@astr1pi.difi.unipi.it}

\and

\author{Robert F. Stellingwerf}
\affil{Los Alamos National Laboratory, X-1 Division, MS F645, Los Alamos,
NM 87545, USA; rfs@demos.lanl.gov}

\altaffiltext{1}{Dipartimento di Fisica, Universit\`a de L'Aquila, Via Vetoio, 67100 L'Aquila, Italy}

\altaffiltext{2}{Osservatorio Astronomico di Teramo, Via M. Maggini, 64100 Teramo, Italy}

\pagebreak
\begin{abstract}

\noindent
We investigate the modal stability of stellar models at masses
and luminosity levels corresponding to post main sequence 
luminous \ds pulsators. The envelope models have been computed at 
fixed mass value (\msun =2.0), luminosity level (\lsun =1.7) and chemical 
composition (Y=0.28, Z=0.02). According to a nonlinear approach to 
radial oscillations the present 
investigation predicts the occurrence of stable \seco pulsators for the 
first time. More in general, we found that when moving inside the 
instability strip from lower to higher
effective temperatures the models show a stable limit cycle in three
different pulsation modes: fundamental, first and second overtone.
The shape of both light and velocity curves are presented and discussed, 
providing a useful tool for the identification
of \seco pulsators among the known groups of radially pulsating stars.   

\noindent 
Comparison with observations shows that our nonlinear, nonlocal and 
time-dependent convective models provide light curves in agreement with 
observed values, suggesting that  second overtone pulsators have 
already been observed though  misclassified as fundamental pulsators.
In a limited region of the instability strip we also found
some models presenting mixed mode features, i.e. radial pulsators 
which show a stable limit cycle in more than one pulsational mode.  

\noindent 
The period ratios of mixed mode pulsators obtained by perturbing the first 
and the second overtone radial eigenfunctions are in agreement with 
observative values. This result is a crucial point for understanding the 
pulsation properties of \ds stars since it provides a sound 
evidence that these variables during their evolution off the main sequence 
are pure or mixed mode radial pulsators. 
Finally, the physical structure and the dynamical properties of \seco 
pulsators are discussed in detail.
The role played by the nodal lines in the destabilization of second overtone 
pulsators is also pointed out. 
\end{abstract}

\noindent
{\em Subject headings:} convection -- $\delta$ Scuti: oscillations -- 
Galaxy: stellar content -- hydrodynamics -- stars: evolution -- 
stars: fundamental parameters

\pagebreak
\section{INTRODUCTION}

\noindent 
The occurrence of variable stars radially pulsating in the
fundamental and/or the first overtone modes is a well
known and well established observational evidence based on   
both Population I and Population II variables. The possible 
occurrence of stars pulsating in the second overtone (SO) is a still 
much debated argument.
For a long time the observational scenario concerning such an occurrence 
has been limited to the suggestions of several authors 
(van Albada \& Baker 1973; Demers \& Wehlau 1977; Clement, Dickens 
\& Bingham 1979; Nemec, Wehlau \& Mendes de Oliveira 1988 and references 
therein) that some globular clusters RRc variables (first overtone) 
characterized by 
very short periods (P$\sim$ 0.3 d) and small pulsational amplitudes could 
be good \so candidates. Only in recent times the large and homogeneous 
photometric databases collected by the MACHO project for radial pulsators 
in the Large Magellanic Cloud (LMC) have brought out for the first time a 
sound evidence of 
the occurrence of SO pulsators in classical Cepheids (Alcock et al. 1995). 
On the basis of the same extensive photometric survey, from 
the period distribution of about 8000 RR Lyrae variables, it has 
been also suggested (Alcock et al. 1996) that a peculiar peak located 
at P=0.28 d could be due to low mass \so variables.

\noindent 
From a theoretical standpoint, the limit cycle stability of \seco 
pulsators in population I and population II variable stars has been an 
odd problem for a long time. As a matter of fact, still at present we lack 
either firm theoretical predictions concerning the existence of \so 
pulsators, or convincing insights on the 
physical mechanisms which could govern their approach to mode stability.  
Classical nonlinear analyses of pulsation properties 
and modal stability of population II low mass variables (Christy 1966; 
Stellingwerf 1975)  never produced any firm evidence of unstable \so 
pulsators. By using a simple nonlinear and nonadiabatic 
one-zone model Stellingwerf, Gautschy \& Dickens (1987, hereinafter 
referred to as SGD) succeeded in producing an unstable \so 
pulsator. However, even though 
one-zone pulsating models can be of some help to disclose the 
main processes which constrain the physical structure of stellar 
envelopes, unfortunately this approach cannot supply definitive 
conclusions on the existence of these variables since these models 
do not take into account the dynamical effects of the regions which 
operate the damping of the pulsation.  
An unstable SO model at limiting amplitude was constructed for RR 
variables  by Stothers (1987) on the basis of a nonlinear radiative approach. 
However, the light curve of his model shows a peculiar feature, 
i.e. a deep splitting just after the maximum in luminosity, 
which is not observed in any known group of radial pulsators and
in the light curve predicted by SGD. Moreover, no evidence of unstable 
SOs was found in the homogeneous and systematic survey of 
nonlinear, nonlocal and time-dependent convective models provided by 
Bono \& Stellingwerf (1994 hereinafter referred to as BS) and by 
Bono et al. (1996a hereinafter referred to as BCCM). 

\noindent
According to the previous discussion,  no \so pulsators appear to be 
foreseen in old, low mass variable stars. In this context, it is worth 
underlining   that periods of RRc in globular clusters are not in conflict 
with first overtone expectations, and their small amplitudes could 
be due to the sudden decrease of the first overtone amplitudes 
near the first overtone blue boundary (BCCM). Moreover, the peak in the 
period distribution at P=0.28 d of RR Lyrae in LMC can be taken as a 
compelling evidence of a moderately metal rich population, with periods 
which overlap the characteristic short periods already found both in the 
Galactic bulge and in the metal rich field RR Lyrae. The reader interested 
to a detailed discussion on the evolutionary and pulsational properties
of metal rich RR Lyrae variables is referred to Bono et al. (1996b). 

\noindent
As far as more massive stars are concerned, it has been often 
suggested that the Hertzsprung progression of the bump in the light 
curve of classical Cepheids could be closely correlated to a {\em spatial 
resonance} between the \so and the fundamental mode (Simon \& Schmidt 1976; 
Buchler et al. 1996). However, 
present nonlinear calculations devoted to the modal stability of 
these variables did not supply any firm conclusion on the existence 
of \so pulsators. Although further nonlinear investigations have 
not been so far 
undertaken to properly tackle the problem, it is worth mentioning
that linear nonadiabatic models provided by Baker (1965) for the  
mass value \msun=1.0 disclosed a \so instability over a wide range 
of luminosities and temperatures. 
Moreover, Deupree \& Hodson (1977), in their study on the instability 
strip of Anomalous Cepheids, gave an evaluation of the \so linear 
blue edge for even larger masses ($1.0 \leq $ \msun $\leq 2.0$). 
However, the quoted authors did not come to a sound conclusion since 
the numerical difficulties caused by the increased efficiency of the 
convective overshooting prevented them from evaluating the nonlinear 
modal stability of these pulsators. 

\noindent 
An extensive grid of \ds pulsation properties has been recently provided 
by Milligan \& Carson (1992) taking into account a combination of stellar 
evolution models and both linear and nonlinear pulsation models. 
Although in this investigation they carried out a detailed analysis 
of the linear and nonlinear properties of \ds stars, only few models 
have been followed at limiting amplitude due to the small value of the 
growth rates involved. Therefore the ultimate modal stability of the wide 
range of nonlinear models could not be firmly established.  
Let us note that \ds variables are most attractive 
objects among the several groups of variables presently known since 
they also show simultaneous excitation of both radial and nonradial modes.
This peculiar feature is of the utmost importance for properly
addressing several astrophysical questions concerning the physical
mechanisms which rule the driving and the quenching of stellar
oscillations. 

\noindent
The observational panorama of \ds stars is rather complex 
since both the rapidly oscillating roAp magnetic stars (Shibahashi 1987;
Martinez \& Kurtz 1995) and the large amplitude metal poor SX Phoenicis 
stars (Nemec, Linnell Nemec \& Lutz 1994; Nemec et al. 1995) belong to 
this group of variable stars as well. However, new high quality photometric 
data of \ds variables in the Galactic field (Rodriguez et al. 1995),
in the Baade's window (Udalski et al. 1995), in open clusters
(Frandsen et al. 1995) and in globular clusters (Nemec et al. 1994)
have been recently provided and therefore a proper classification of 
these objects based on their evolutionary and pulsational properties
will be soon available.   
Another interesting aspect which makes \ds variables worth being investigated 
is that during their evolution intermediate mass stars cross the instability 
strip in different evolutionary phases. In fact, in contrast with 
canonical Cepheid-like variables which are connected with Helium burning 
evolutionary phases, in this region of the HR diagram the current evolutionary 
theory foresees stars in pre-main-sequence, main sequence and post main 
sequence phases. As a consequence, their pulsational properties can supply 
useful constraints on their evolutionary properties and at the 
same time an independent check of stellar models (Breger 1993,1995). 

\noindent
Linear adiabatic and nonadiabatic models of these objects have been 
computed by Stellingwerf (1978), Andreasen, Hejlesen \& Petersen (1983) and 
Andreasen (1983). Even though these investigations have significantly 
improved our knowledge of the location of the linear blue boundaries of the 
instability strip, the theoretical scenario of \ds stars presents 
several unsettled problems. 
According to this evidence, we decided to extend the investigation on 
modal stability and pulsational properties already provided for low mass 
He burning stars to larger mass values. Canonical evolutionary prescriptions 
indicate that stars with 
masses of the order of (1.5 - 2.0)$M_{\odot}$, unlike low mass stars, 
spend a non negligible portion of their lifetime inside the instability 
strip during their evolution off the Main Sequence. Therefore relatively 
massive stars are expected to generate, during this phase, short period 
radial pulsators. 

\noindent 
In this paper we investigate the nonlinear pulsational properties of this 
kind of variable stars (Eggen 1994; McNamara 1995) by exploring the 
pulsational behavior of a sequence of models as a function of the surface 
effective temperature. For the first time we show that the adopted 
theoretical framework accounts for the modal stability of \so pulsators 
and allows some predictions concerning the full amplitude behavior of 
\so light and velocity curves. In \S 2 we describe the numerical and physical 
assumptions adopted for constructing both linear and nonlinear pulsation
models, whereas in \S 3 we discuss the full amplitude pulsation behavior 
for the first three modes. The shape and the secondary features of both light 
and velocity curves are presented in \S 4 together with a plain comparison
with available photometric data. The physical parameter which rules the 
\so limit cycle stability and their dynamical properties are discussed 
in \S 5. Our conclusions are  presented in \S 6. In this section the 
consequences arising from these new findings and the observative features 
worth being investigated are outlined as well.

\section{STANDARD PULSATION MODELS}

\noindent
The theoretical framework adopted for investigating linear and nonlinear 
pulsation characteristics of radial pulsators have been previously 
described in a series of papers (Stellingwerf 1982; BS and references 
therein). The sequence of models presented in this paper was 
computed at fixed mass
value (\msun =2.0), chemical composition (Y=0.28, Z=0.02) and luminosity 
level (\lsun =1.7) and by exploring a wide range of effective 
temperatures ($7500 \geq T_e \geq 6000$ K).   
The static envelope models were constructed by adopting an optical 
depth of the outermost zone of the order of 0.001, and the inner boundary 
was assumed fixed so that possible destabilization effects due to variations  
in the efficiency of the H shell burning are ignored.  
Moreover, we adopted the OP radiative opacities provided by Seaton et al. 
(1994) for temperatures higher than 10,000 K whereas for lower temperatures 
we adopted  molecular opacities provided by Alexander \& Ferguson (1994). 
The method adopted for handling the opacity tables was already described 
in Bono, Incerpi \& Marconi (1996). 

\noindent
Each envelope model extends from the surface to 20-10\% of the stellar 
radius and the zone closest to the Hydrogen Ionization Region (HIR) was
constrained to $T_{HIR}=1.3 \times 10^4$ K.  
Between the HIR and the surface we inserted 20 zones to ensure a good 
spatial resolution of the outermost regions throughout the pulsation cycle.  
Due to the key role played by spatial resolution for firmly estimating 
both linear and nonlinear modal stability of higher modes, the envelopes 
were discretized by adopting a detailed zoning in mass. The mass 
ratio between consecutive zones ($\theta$) has been assumed equal to 
$\theta$=1.1 for temperatures lower than $6 \times 10^5$ K, whereas 
for higher temperatures it has been set equal to $\theta$=1.2. 
By adopting this type of zoning the ionization 
regions and the opacity bump due to iron are covered with a number of 
zones lying between 100 and 150. 
The fine spatial resolution of the ionization fronts provides, in turn, 
an accurate treatment of both the formation and the propagation of shock 
fronts during the phase interval between the phase of minimum radius and 
the phase of maximum luminosity (for complete details see BS). 
On the basis of these assumptions a typical envelope model is characterized 
by roughly 20-30\% of the total stellar mass and by 150-250 zones.   

\noindent 
To clarify matters concerning the dependence of the 
pulsation behavior on the spatial resolution we have constructed a 
much finer envelope model. This model is located at $T_e=7000$ K and, in 
contrast with the standard sequence of linear nonadiabatic models,  
it was constructed by adopting a smaller mass ratio ($\theta$=1.08)
for the regions located at temperatures lower than $6 \times 10^5$ K. 
Moreover, the inner boundary condition for this model was 
chosen in such a way that the base of the envelope was located below
0.1 of the photospheric radius and its temperature was of the order of
$5-6 \times 10^6$ K. 

\noindent 
As a consequence of these assumptions the detailed model 
presents an increase in both the envelope mass ($M_{env}=0.46 M_T$ 
against $M_{env}=0.21 M_T$) and the number of zones which cover the 
ionization regions (180 against 150). We eventually found that 
for the first three modes the differences between the pulsation 
characteristics of both detailed and standard model are quite negligible.  
In fact, the discrepancies range from $10^{-4}$ to $10^{-3}$ 
for the periods and from $10^{-6}$ to $10^{-4}$ for the growth rates.  
As a result we can assume that the spatial resolution of the standard 
sequence allows a proper treatment of the radial pulsation of \ds stars
during their off main sequence evolution.

\section{APPROACH TO LIMIT CYCLE STABILITY}

\noindent 
According to the usual approach, a sequence of linear nonadiabatic models 
was first constructed for supplying the static structure of the envelope 
to the nonlinear stability analysis. Then the equations governing both the 
dynamical and the convective structures were integrated in time until 
the initial perturbations and the nonlinear fluctuations, which result from 
superposition of higher order modes, settled down (for more detail see 
Bono, Castellani \& Stellingwerf 1995). 
The dynamical behavior of the envelope models was computed for 
the first three modes and the initial velocity profile was obtained 
by perturbing the linear radial eigenfunctions with a constant 
velocity amplitude of 20 km$s^{-1}$ which causes a global expansion 
of the envelope. 

\noindent 
As is well known, the linear nonadiabatic \ds models present very small
growth rates and therefore, before radial motions approach the nonlinear 
limit cycle stability, it is necessary to carry out extensive calculations. 
In fact, the long-term stability of a particular mode, due to the mixture
of both periodic and nonperiodic motions characterized by very small growing 
and/or decaying rates, cannot be easily assessed at small amplitude.  
As a consequence, in order to find out the possible appearance of a mode 
switching or of a mixture of modes we evaluated the asymptotic behavior 
of each mode by performing very long runs.
This approach leads to an integration of the governing equations for a 
number of periods which ranges from 5,000 to 50,000 for some peculiar 
cases. The integration is generally stopped when the nonlinear total
work is vanishing and the pulsation amplitudes present over two consecutive 
periods a periodic similarity of the order of or lower than 
$10^{-(2 \div 3)}$.  

\noindent 
Since this is the first time that hydrodynamic calculations are performed
over such a long time interval, Figures 1, 2 and 3 show the time behavior of
period, velocity and magnitude for three cases characterized by a different
approach to nonlinear limit cycle stability. In particular, Fig. 1 shows the 
variation 
of the quoted quantities for a single pure \so model, whereas Fig. 2 shows 
that radial motions at $t\approx 7$ yrs experience a mode switching from 
the first overtone to the \so. Figure 3 finally presents the limiting 
amplitude behavior of a case which presents a permanent mixture of different 
radial modes. 

\noindent 
As a result of the modal stability analysis, we found stable nonlinear limit 
cycles in the fundamental, in the first overtone and, for the first time, 
in the second overtone when the effective temperature is increased.  
Even though so far the location inside the instability strip of \ds stars 
characterized by different pulsation modes has not been firmly 
established, the previous finding confirms the distribution originally 
suggested by Breger \& Bregman (1975). 
In fact, by assuming that \ds variables are radial pulsators, these authors 
found that observed second and first overtone variables were located at 
effective temperatures higher than the fundamental ones. 

\noindent 
In Table 1 are listed selected observational parameters for the sequence 
of \ds models. As a first result,
data in Table 1 show that theoretical periods appear in general 
agreement with the observed range of \ds values. 
It is worth underlining that the effective temperature of the fundamental 
red edge should be considered an upper limit. As a matter of fact, even 
though the model located at 6300 K after 23,000 periods presents 
both a constant negative value in the total work term and a very low 
pulsational amplitude ($\Delta M_{bol} \approx 4-6 \times 10^{-3}$ mag), 
this region of the instability strip, due to the slow approach to limit 
cycle stability, should be investigated in more detail before firmly 
constraining the location of the red edge. 

\noindent 
In order to disclose the main features of the modal behavior in \ds 
stars, Figure 4 shows the bolometric light curves and the surface 
radial velocities of \so pulsators; Figure 5 shows the same quantities 
but they are referred to selected first overtone (solid lines) and 
fundamental (dashed lines) pulsators. 
Figure 6 shows the light and radial velocity curves 
of mixed mode pulsators, i.e. of models which present a permanent mixture 
of different radial modes at limiting amplitude. 

\noindent 
Inspection of light curves discloses the surprising evidence that
the shape of \so light curves -sudden increase in the rising branch and slow 
decrease in the decreasing branch- 
closely resembles  canonical fundamental mode rather than first overtone 
RR Lyrae pulsators. This finding confirms the original prediction 
concerning the shape of \so light curves made by SGD. 
Moreover, we find that moving from \so to lower 
pulsational modes the amplitudes progressively decrease and the shape 
of the light curves becomes more sinusoidal. It is worth noting that the 
pulsation amplitudes of RR Lyrae variables, which are located in the same 
region of the instability strip, present an opposite trend. 
In fact, for this group of pulsators the RRab variables (fundamental) 
show the largest amplitudes. These theoretical prescriptions can be 
usefully compared with the observational scenario recently discussed  
by McNamara (1995 and references therein). 

\noindent
According to this author, \ds stars on the basis of their luminosity  
amplitude can be empirically divided into two groups.  
The light curves of stars with larger amplitudes appear to be 
asymmetrical whereas the light curves for lower amplitudes
tend to be much more symmetrical. However, in the above paper McNamara also 
suggests that for low amplitude variables, which are poorly sampled, 
it is often difficult to determine whether the light curves are symmetric 
or asymmetric. Therefore, for light curves which are only partially 
covered by photometric data, several cases of probable asymmetry are brought 
forward. By analogy with the behavior of RR Lyrae variables, McNamara (1995) 
assumes that stars with asymmetric, large amplitude
light curves are fundamental pulsators whereas symmetric, low amplitude
light curves belong to first overtone pulsators.  
The comparison of similar empirical prescriptions with the current  
theoretical scenario shows a convincing degree of agreement.
However, theory now tells us that asymmetric, large amplitude pulsators
are good \so candidates, whereas low amplitude pulsators could be a
mixture of fundamental and first overtones.

\noindent
To go further on with this comparison, let us refer to the sample of variable 
stars recently collected by the OGLE collaboration (Udalski et al. 1995 and 
references therein) as the result of their search for evidence of 
microlensing in the bulge of the Galaxy. 
Inspection of photometric data connected with short period variables 
discloses that the observed light curves can be arranged in three typical 
classes, as shown in Fig. 7, with class "A" representing McNamara large 
amplitude pulsators and classes "B" and "C" the small amplitudes ones.
For a meaningful comparison between theory and observation, the 
bolometric light curves have been transformed into the I band according 
to Kurucz's (1992) atmosphere models. Figures 8 and 9 show the light 
curves for single mode pulsators. 
Due to the magnitude scale adopted for plotting data in Fig. 9, the light 
curves of fundamental pulsators (dashed lines) seem almost perfectly 
sinusoidal. However, even though the luminosity variations throughout 
the cycle are quite smooth, a bump appears before the phase of minimum 
radius. Taking into account that we explored only one mass value and 
only one luminosity level, the comparison should be considered more 
than satisfactory.

\section{ LIGHT CURVE MORPHOLOGY}

\noindent
Available observational data hardly allow to detect minor details in the 
light curves. However, the quality of both spectroscopic and 
photometric data is rapidly improving (see for example Milone, Wilson 
\& Fry 1994; Breger et al. 1995 and references therein) and therefore 
in this section we discuss even minor features of theoretical light curves  
in order to underline the theoretical predictions worth being 
investigated with the required accuracy. As a first point, let
us notice that the light and velocity curves of \so pulsators 
present two further relevant distinctive features: 

\noindent
1) like canonical first overtone RR Lyrae variables, the bump does not 
appear along the decreasing branch of the light curves and, moving from 
higher to lower effective temperatures the dip becomes more and more 
evident along the rising branch; 

\noindent
2) moving from the blue to the red boundary of the SO instability region  
the velocity curves show smooth variations but at phases 0.2-0.3 
a bump appears due to the propagation of an outgoing shock. 

\noindent
As a second point, we find that the shape of first overtone light curves 
presents some features which allow a careful distinction 
between different radial modes. In fact for these models the dip is the 
main maximum, whereas the "true" maximum takes place along the decreasing 
branch. Moreover, the first overtone light curves show that the bump 
appears along the increasing branch and that just before the phase of 
minimum radius they also display a short stillstand phase 
($\phi \approx 0.45$). 

\noindent
The scenario concerning the pulsation characteristics of \ds stars can 
be now nicely completed by the models which are simultaneously
excited in two or more radial modes. Figure 6 shows a collection of 
light and velocity curves of mixed mode pulsators, a glance to these curves
brings out both the expected fluctuation of the pulsation amplitudes between 
consecutive cycles and the appearance of the secondary features which 
characterize the first three lower single mode pulsators. 

\noindent
A thorough comparison with observed period ratios of \ds variables is beyond 
the scope of the present study since the "true" period ratios should be 
evaluated through a Fourier decomposition of the theoretical light curves. 
Moreover, for properly constraining the stellar masses and the luminosity 
levels of these objects by means of the Petersen diagram 
($P_1$/$P_0$ vs. $P_0$) an extensive set of nonlinear models computed for 
different assumptions on astrophysical parameters is necessary 
(Bono et al. 1996c). 

\noindent
Nevertheless, since period ratios of double mode pulsators can 
provide valuable clues on several astrophysical problems involving both the 
evolutionary and the pulsational properties of these objects, we constructed 
a new sequence of detailed, linear, radiative, nonadiabatic models by adopting  
the assumptions already discussed at the end of section 2. This analysis
has been undertaken only for supplying a preliminary but meaningful theoretical 
guess concerning the location of this group of variables inside the instability 
strip. 

\noindent
At first it is important to note that the linear periods  
and the related period ratios listed in Table 2 are, within the estimated 
uncertainties, in agreement with 
observed values (see for example data on double mode \ds stars collected by 
Andreasen 1983 and by Petersen 1990). Indeed the observed period ratio between
first overtone and fundamental pulsators ranges, for large amplitude \ds 
stars, from 0.760 to 0.780, whereas the period ratio between second overtone 
and first overtone is approximatively of the order of 0.800.  
This agreement is a remarkable result since the period ratios predicted by 
pulsation models are the most important 
observable adopted for finding out whether the pulsation is "driven" by 
radial or by nonradial modes. As a consequence, this finding provides sound
evidence that these variables are mixed mode radial pulsators (Breger 1979). 

\noindent
Moreover, previous linear and nonlinear results suggest that, due to the 
appearance of three different modal stabilities inside the instability 
strip, double mode pulsators belonging to this group of variables are 
located close 
to the fundamental blue edge when they are a mixture of fundamental and 
first overtone and close to the first overtone blue edge when they are 
a mixture of first and second overtones. 
A detailed investigation on the dependence of this peculiar occurrence among  
radial pulsators together with a straightforward analysis of the envelope 
structure will be discussed in a forthcoming paper (Bono et al. 1996d).

\section {SECOND OVERTONE INSTABILITY}

\noindent
In order to properly identify the regions of the stellar envelope which  
{\em drive} or {\em damp} the pulsation instability of \so pulsators,  
Fig. 10 shows the nonlinear differential work integrals versus 
the logarithm of the external mass for a model located close to 
the \so blue edge. In this plane the positive areas denote driving
regions (growing oscillations) whereas the negative areas damping
regions (quenching oscillations). The total work curve shows quite 
clearly the two driving sources due to the hydrogen and helium 
ionization zones as well as the radiative damping due to the inner 
regions. Unlike in canonical cluster variables, the second helium 
ionization zone provides a stronger destabilization if compared  
with the HIR. This effect is mainly due 
to the increase in effective temperature which causes a shift of 
the HIR toward the surface and therefore a decrease of the mass 
which lies above these layers. However, the total work plotted in Fig. 10 
clearly shows that this element, in contrast with previous 
qualitative arguments, provides a substantial amount of driving to 
the pulsation instability of \ds stars. 
For the reasons previously discussed all other nonlinear work 
terms supply a negligible damping effect on the pulsation. 

\noindent
Nevertheless the physical parameter which rules the \so instability 
is the location of the nodes 
\footnote{In the case of a vibrating membrane a {\em node} is the 
point where an eigenfunction vanishes or attains a sharp minimum and 
its phase changes by almost $\pi$ radians.} 
inside the envelope. In fact, 
the nodes of temperature, luminosity and radius fall within the region of
radiative damping. As a consequence, the amount of damping is strongly
reduced and the destabilization of the ionization regions pumps up 
the pulsation amplitude. In particular, it is worth emphasizing  
that among cluster variables the outermost node of radius is located 
quite close to the helium driving region (Stellingwerf 1990). 
In this context the node of radius has an opposite effect since it 
reduces the amount of driving and consequently quenches the oscillations. 

\noindent
On the basis of a well-known theorem concerning the Sturm-Liouville 
eigenvalue problem, a \so eigenfunction should split its domain into 
two parts by means of its {\em nodal lines} (Courant \& Hilbert 1989). 
Even though during a full pulsation cycle the radial motions never 
exactly approach the initial equilibrium configuration, the velocity 
curves plotted in Fig. 11 undoubtedly show two different 
subdomains characterized by opposite radial motions.  
In order to disclose the dynamical structure of the model previously 
discussed, Fig. 11 shows the radial displacements of the whole 
envelope over two consecutive periods. Dots and pluses mark the phases 
during which each zone is contracting or expanding respectively. 
It is easy to ascertain from the above figure that at a fixed pulsation 
phase the envelope is divided into two different regions and the velocity 
curves are exactly 180$^{\circ}$ out of phase at their boundaries. 
In particular, the layers located close to the two boundaries define two 
zones which virtually remain at rest throughout the pulsation 
cycle and in which the radial velocity changes its sign abruptly.  

\section{CONCLUSIONS}

\noindent
In this paper we report the first theoretical evidence for the
occurrence of pulsators that when moving inside the instability strip 
from lower to higher effective 
temperatures show three different stable pulsation modes, namely  
fundamental, first and second overtone. 
We found that the predicted features of the light curves  
appear in general agreement with observational constraints
concerning \ds variables, suggesting that \so pulsators have
been already observed and pointing out, at the same time, a revised 
approach to observational evidences for different pulsation modes.
Moreover, we show that the location of radius, luminosity and temperature
nodes in the damping region of the stellar envelope is the main 
physical parameter which governs the limit cycle stability of \so 
pulsators. This result strongly supports the discussion given in the 
introduction to this paper about the absence of \so pulsators in low mass
variable stars. 

\noindent
The satisfactory agreement between the computed period ratios and the 
observed ones casts new light on the problem of limit cycle stability 
of mixed mode variables belonging to population II stars and provides 
a valuable piece of information for accounting for the pulsation 
properties of \ds stars. At the same time there is a growing strong 
evidence that these stars during their evolution off the main sequence 
are pure or mixed mode radial pulsators. 

\noindent
Before being able to provide a sound comparison with available photometric 
data on \ds variables, the present sequence of nonlinear models should be 
extended to both lower luminosity levels and to smaller mass values.
However, even though the computed models cover a restricted range of 
effective temperatures, the theoretical framework currently adopted 
accounts for both pure and mixed mode radial pulsators. This new scenario 
presents several interesting features since both the modal stability and 
the pulsational behavior have been investigated in a homogeneous 
physical context without invoking unpleasant {\em ad hoc} physical 
mechanisms and/or peculiar characteristics.  

\noindent
Plenty new high quality photometric data on \ds stars will be soon 
available as a by-product of the international projects involved in the 
search for microlensing events and therefore new sequences of nonlinear \ds
models at full amplitude, albeit such analysis places nontrivial
computational efforts, are necessary to firmly accomplish the pulsation 
properties of these objects. 

\noindent
Finally we suggest that short period RR Lyrae-like pulsators found in 
the Galactic bulge as well as in dwarf spheroidals like Carina and 
Sagittarius should be regarded as an evidence of relatively massive 
stars, a witness of the efficiency of star formation until relatively 
recent times. On the other hand this finding stresses once again the key 
role played by variable stars as tracers of stellar populations which 
experienced different dynamical and/or chemical evolutions. 

\noindent
It is a pleasure to thank F. Pasian and R. Smareglia for their kind 
and useful sharing of computing facilities which made this investigation 
possible. We are grateful to J. Nemec for his clarifying suggestions as 
referee on an early draft of this paper. We also benefit from the use of 
the SIMBAD data retrieval system (Astronomical Data Center, Strasbourg, 
France). This work was partially supported by MURST, CNR-GNA and ASI.

\pagebreak

\pagebreak
\section{FIGURE CAPTIONS}

\vspace*{3mm} \noindent {\bf Fig. 1.} - Time behavior of period ({\em top 
panel}), velocity ({\em middle panel}) and bolometric amplitude 
($\Delta M = \Delta M_{bol}(max) - \Delta M_{bol}(min)$; ({\em bottom panel}) 
for a single pure second overtone model located at Te=7150 K.  
The static structure is forced out of its equilibrium status by perturbing 
the \so radial eigenfunction with a constant velocity amplitude of 
20 km$s^{-1}$.  

\vspace*{3mm} \noindent {\bf Fig. 2.} - Same as Fig. 1, but referred to a 
case, located at Te=7000 K, which shows a mode switching from the first 
overtone to the SO. The arrow marks the time at which the mode switching 
takes place.   

\vspace*{3mm} \noindent {\bf Fig. 3.} - Same as Fig. 1, but referred to a 
case, located at Te=7000 K, which shows a permanent mixture of different 
radial modes at limiting amplitude. This model was initiated by perturbing
the first overtone radial eigenfunction. 

\vspace*{3mm} \noindent {\bf Fig. 4.} - {\em Top:} Second overtone bolometric  
magnitude for two consecutive periods. The magnitude 
scale refers to the top light curve; the other curves are artificially 
shifted by 1 mag. The labels show the periods (days) for the selected 
\so models.  
{\em Bottom:} Second overtone radial velocity curves. The velocity scale 
refers to the top velocity curve; the other curves are artificially shifted 
by -65 km$s^{-1}$. The labels show for the selected \so models the 
effective temperatures (K). Diamonds mark the phase of minimum radius. 

\vspace*{3mm} \noindent {\bf Fig. 5.} - Same as Fig. 4, but the solid lines
are referred to pure first overtone pulsators and the dashed lines to 
fundamental pulsators. 

\vspace*{3mm} \noindent {\bf Fig. 6.} - {\em Top:} Time behavior of mixed 
mode bolometric magnitudes over two consecutive periods. In this context
the period is defined as the time interval between two consecutive 
absolute maxima in the surface radius. 
The magnitude scale refers to the top light curve; the other curves are 
artificially shifted by 1.5 mag. {\em Bottom:} Mixed mode radial velocity 
curves over the same time interval adopted for the light curves. 
The velocity scale refers to the top velocity curve; the other curves are 
artificially shifted by -85 km$s^{-1}$.
The solid curve is referred to a model that was initiated by perturbing 
the fundamental radial eigenfunction, whereas the dashed curves are 
referred to models that were initiated by perturbing the first overtone 
radial eigenfunction. 

\vspace*{3mm} \noindent {\bf Fig. 7.} - Light curves of \ds variables 
selected from the OGLE photometric database. The variables were chosen
in such a way that the top curve can be considered a template for 
second overtone pulsators (class "A"), whereas the middle curve for 
first overtone pulsators (class "B") and the bottom one for fundamental 
pulsators (class "C").  

\vspace*{3mm} \noindent {\bf Fig. 8.} - Second overtone light
curves for two consecutive periods. The luminosity curves have been 
transformed into the I band by adopting the static atmosphere models  
provided by Kurucz (1992). The magnitude scale refers to the top light 
curve; the other curves are artificially shifted by 1 mag. Diamonds 
mark the phase of minimum radius. 

\vspace*{3mm} \noindent {\bf Fig. 9.} - Same as Fig. 8, but the solid 
lines are referred to pure first overtone pulsators, whereas the dashed 
lines to pure fundamental pulsators. 

\vspace*{3mm} \noindent {\bf Fig. 10.} - Nonlinear work integrals for a 
\seco vs. the logarithm of the external mass, surface at right. 
The solid line shows the total work, the dashed line the total turbulent 
work (i.e. turbulent pressure plus eddy viscosity works) and the 
dashed-dotted line the artificial viscosity work. The different symbols mark, 
according to linear eigenfunctions, the location of the nodes in radius, 
temperature and luminosity. The hydrogen and helium ionization regions 
are also indicated. 

\vspace*{3mm} \noindent {\bf Fig. 11.} - Second overtone velocity 
variations of the whole envelope for two consecutive periods at 
limiting amplitude. Starting from the surface zone (top), the inner 
regions are plotted every four zones. 
For each zone dots and pluses mark the phases during which the radial 
velocity assumes negative (contraction) or positive (expansion) values 
respectively. The velocity scale is in arbitrary units since moving 
from the surface to the innermost regions each curve  
has been both shifted and enhanced artificially by increasing factors  
in order to make the overall dynamical structure of the envelope as 
plain as possible. The arrows display the location of the nodal lines.


\clearpage 
\begin{deluxetable}{ccccccccc}
\footnotesize  
\tablecaption{Nonlinear Survey: Observables}\label{tbl-1} 
\tablehead{
\colhead{$T_e$\tablenotemark{a}}& 
\colhead{Period\tablenotemark{b}}& 
\colhead{$\Delta R/R_{ph}$\tablenotemark{c}}& 
\colhead{$\Delta u$\tablenotemark{d}}& 
\colhead{$\Delta M_{bol}$\tablenotemark{e}}& 
\colhead{$\Delta {\log g_s}$\tablenotemark{f}}& 
\colhead{$\Delta {\log g_{eff}}$\tablenotemark{g}}&  
\colhead{$\Delta T$\tablenotemark{h}}& 
\colhead{$\Delta T_e$\tablenotemark{i}} }
\startdata
        \multicolumn{9}{c}{Second Overtone}\nl  
7300 ($+$)\tablenotemark{j}& .1305 & .051 & 96.85 & 1.079& .04 & .95 & 1650 & 2050 \nl
7150 ($+$)& .1390 & .054 & 97.93 & .965 & .05 & .99 & 1450 & 1800 \nl     
7000 ($+$)& .1484 & .054 & 95.73 & .716 & .05 & 1.04& 1050 & 1300 \nl  
6850 ($+$)& .1581 & .048 & 82.66 & .506 & .04 & .94 & 750  & 950  \nl
6700 ($\;\ast\;$)& .2409 & .017 & 24.01 & .105 & .03 & .49 & 400  & 500  \nl
       &       &      &       &      &     &     &      & \vspace*{-4.0mm}\nl
       \multicolumn{9}{c}{First Overtone} \vspace*{-4.0mm}\nl 
       &       &      &       &      &     &     &      & \nl
7000 ($\;\ast\;$) & .1481 & .054 & 95.31 & .713 & .05 & 1.04& 1000 & 1200 \nl 
6850 (\#) & .1994 & .064 & 86.00 & .648 & .06 & 1.09& 1650 & 2000 \nl
6700 ($+$)& .2133 & .074 & 99.44 & .596 & .06 & .86 & 900 & 1100 \nl     
6600 ($+$)& .2230 & .066 & 83.56 & .509 & .06 & .75 & 750 & 950 \nl  
       &       &      &       &      &     &     &      & \vspace*{-4.0mm}\nl
	\multicolumn{9}{c}{Fundamental} \vspace*{-4.0mm}\nl 
       &       &      &       &      &     &     &      & \nl
7000 (\#) & .2459 & .045 & 57.89 & .580 & .04 & .46 & 800 & 1000 \nl
6700 (\#) & .2745 & .039 & 40.63 & .216 & .02 & .28 & 300 & 400 \nl     
6500 ($+$)& .3115 & .033 & 32.98 & .135 & .03 & .21 & 200 & 250 \nl  
6400 ($+$)& .3264 & .023 & 20.55 & .076 & .02 & .13 & 150 & 150 \nl 
\enddata
\tablenotetext{a}{ Effective temperature (K). 
\hspace*{0.5mm} $^b$ Nonlinear period (days).
\hspace*{0.5mm} $^c$ Fractional radial oscillation, where $R_{ph}$ is the photospheric radius. 
\hspace*{0.5mm} $^d$ Radial velocity amplitude (Km s$^{-1}$). 
\hspace*{0.5mm} $^e$ Bolometric amplitude (mag.). 
\hspace*{0.5mm} $^f$ Logarithmic variation of the surface static gravity (cm s$^{-1}$).  
\hspace*{0.5mm} $^g$ Logarithmic variation of the surface effective gravity (cm s$^{-1}$).
\hspace*{0.5mm} $^h$ Variation of the surface temperature (K). 
\hspace*{0.5mm} $^i$ Variation of the effective temperature (K). 
\hspace*{0.5mm} $^j$ The symbols enclosed in parentheses are referred to the 
modal stability. Models characterized by a stable limit cycle are denoted
by a plus ($+$), whereas models which show a mode switch are denoted by an 
asterisk ($\ast$). Models which present a permanent mixture of different 
radial modes at full amplitude are denoted by a hash $(\#)$.} 
\end{deluxetable}

\clearpage 
\begin{deluxetable}{cccccccccc}
\footnotesize  
\tablecaption{Linear Nonadiabatic Periods}
\label{tbl-1} 
\tablehead{
\colhead{$T_e$\tablenotemark{a}}& 
\colhead{$P(F)$\tablenotemark{b}}& 
\colhead{$P(FO)$\tablenotemark{c}}& 
\colhead{$P(SO)$\tablenotemark{d}}& 
\colhead{$Q(F)$\tablenotemark{e}}& 
\colhead{$Q(FO)$\tablenotemark{f}}& 
\colhead{$Q(SO)$\tablenotemark{g}}&  
\colhead{${P_{FO}}/{P_{F}}$\tablenotemark{h}}&
\colhead{${P_{SO}}/{P_{FO}}$\tablenotemark{i}}& 
\colhead{${P_{SO}}/{P_{F}}$\tablenotemark{j}} }
\startdata
7400 & .2067& .1589& .1263& .0330& .0254&  .0202& .769& .795& .611 \nl
7300 & .2158& .1656& .1316& .0331& .0254&  .0202& .767& .795& .610 \nl
7150 & .2305& .1764& .1402& .0332& .0254&  .0202& .765& .795& .608 \nl     
7000 & .2465& .1880& .1496& .0334& .0254&  .0202& .763& .796& .607 \nl  
6850 & .2641& .2007& .1598& .0335& .0255&  .0203& .760& .796& .605 \nl 
6700 & .2835& .2147& .1711& .0336& .0255&  .0203& .757& .797& .604 \nl
6500 & .3124& .2353& .1876& .0338& .0255&  .0203& .753& .797& .601 \nl   
6400 & .3284& .2467& .1969& .0340& .0255&  .0204& .751& .798& .600 \nl
6300 & .3454& .2587& .2066& .0341& .0255&  .0204& .749& .799& .598 \nl 
\enddata
\tablenotetext{a}{ Effective temperature (K). 
\hspace*{0.5mm} $^b$ Fundamental period (days).  
\hspace*{0.5mm} $^c$ First overtone period (days).  
\hspace*{0.5mm} $^d$ Second overtone period (days).  
\hspace*{0.5mm} $^e$ Fundamental pulsation constant (days).  
\hspace*{0.5mm} $^f$ First overtone pulsation constant (days).  
\hspace*{0.5mm} $^g$ Second overtone pulsation constant (days).  
\hspace*{0.5mm} $^h$ Period ratio between first overtone and fundamental.  
\hspace*{0.5mm} $^i$ Period ratio between second overtone and first overtone.  
\hspace*{0.5mm} $^j$ Period ratio between second overtone and fundamental.} 
\end{deluxetable}


\begin{thebibliography}{}
\bibitem []{} Alcock, C., et al. 1995, AJ, 109, 1654  
\bibitem []{} Alcock, C., et al. 1995, AJ, 111, 1146  
\bibitem []{} Alexander, D. R., \& Ferguson, J. W. 1994, ApJ, 437, 879 
\bibitem []{} Andreasen, G. K., 1983, A\&A, 121, 250 
\bibitem []{} Andreasen, G. K., Hejlesen, P. M., \& Petersen, J. O. 1983, A\&A, 121, 241 
\bibitem []{} Baker, N. 1965, Kl. Veroff. Remeis-Sternw. Bamberg, 4, 122  
\bibitem []{} Bono, G., Caputo, F., Cassisi, S., Castellani, V., Marconi, M., 
\& Stellingwerf, R. F. 1996b, ApJS, submitted      
\bibitem []{} Bono, G., Caputo, F., Castellani, V., \& Marconi, M. 1996a, A\&AS, accepted (BCCM) 
\bibitem []{} Bono, G., Caputo, F., Castellani, V., \& Marconi, M. 1996c, ApJ,   submitted 
\bibitem []{} Bono, G., Caputo, F., Castellani, V., \& Marconi, M. 1996d, in preparation 
\bibitem []{} Bono, G., Castellani, V., \& Stellingwerf, R. F. 1995, ApJ, 445, L145  
\bibitem []{} Bono, G., Incerpi, R., \& Marconi, M. 1996, ApJ Letters, accepted  
\bibitem []{} Bono, G., \& Stellingwerf, R. F. 1994, ApJS, 93, 233 (BS)   
\bibitem []{} Breger, M. 1979, PASP, 91, 5    
\bibitem []{} Breger, M. 1993, Confrontation Between Stellar Pulsation and 
Stellar Evolution, eds. C. Cacciari \& G. Clementini (San Francisco: ASP 83),
263 
\bibitem []{} Breger, M. 1995, in IAU Colloq. 155, Astrophysical Applications 
of Stellar Pulsation, eds. R.S. Stobie \& P.A. Whitelock (San Francisco: 
ASP 83), 70  
\bibitem []{} Breger, M., \& Bregman, J. N. 1975, ApJ, 200, 343 
\bibitem []{} Breger, M., et al. 1995, A\&A, 297, 473 
\bibitem []{} Buchler, J. R., Kollath, Z., Beaulieu, J. P., \& Goupil, M. J. 
1996, ApJ, 462, L83    
\bibitem []{} Christy, R. F. 1966, ApJ, 144, 108  
\bibitem []{} Clement, C. C., Dickens, R. J., \& Bingham, E. E. 1979, AJ, 84, 217
\bibitem []{} Courant, R., \& Hilbert, D. 1989, in Methods of Mathematical 
Physics, (New York: Wiley), 454     
\bibitem []{} Demers, S., \& Wehlau, A. 1977, AJ, 82, 620  
\bibitem []{} Deupree, R. G., \& Hodson, S. W. 1977, ApJ, 218, 654  
\bibitem []{} Eggen, O. J. 1994, AJ, 107, 2131 
\bibitem []{} Frandsen, S., Viskum, M., Hernandez, M. M., \& Belmonte, J. A. 
1995, in IAU Colloq. 155, Astrophysical Applications of Stellar Pulsation,
eds. R.S. Stobie \& P.A. Whitelock (San Francisco: ASP 83), 327
\bibitem []{} Kurucz, R. L. 1992, in IAU Symp. 149, The Stellar Populations of 
Galaxies, eds. B. Barbuy \& A. Renzini, (Dordrecht: Kluwer), 225  
\bibitem []{} Martinez, P., \& Kurtz, D. W. 1995, in IAU Colloq. 155, 
Astrophysical Applications of Stellar Pulsation, eds. R.S. Stobie \& P.A. 
Whitelock (San Francisco: ASP 83), 58  
\bibitem []{} McNamara, D. H. 1995, AJ, 109, 1751   
\bibitem []{} Milligan, H., \& Carson, T. R. 1992, Ap\&SS, 189, 181  
\bibitem []{} Milone, E. F., Wilson, W. J. F., \& Fry, D. J. I. 1994, PASP, 106, 1120   
\bibitem []{} Nemec, J. M., Linnell Nemec, A. F., \& Lutz, T. E. 1994, AJ, 108, 222 
\bibitem []{} Nemec, J. M., Mateo, M., Burke, M., \& Olszewski, E. W. 1995, AJ, 110, 1186  
\bibitem []{} Nemec, J. M., Wehlau, A. \& Mendes de Oliveira, C. 1988, AJ, 96, 528 
\bibitem []{} Petersen, J. O. 1990, A\&A, 238, 160  
\item Rodriguez, E., Rolland, A., Lopez de Coca, P. \& Martin, S. 1995, in
IAU Colloq. 155, Astrophysical Applications of Stellar Pulsation,
eds. R.S. Stobie \& P.A. Whitelock (San Francisco: ASP 83), 329 
\bibitem []{} Seaton, M. J., Yuan, Y., Mihalas, D. \& Pradhan, A. K. 1994,
MNRAS, 266, 805 
\bibitem []{} Shibahashi, H. 1987, in Stellar Pulsation, eds. A. Cox, W.M. 
Sparks \& S.G. Starrfield (Lecture Notes in Physics 274, Berlin: 
Springer-Verlag), 112 
\bibitem []{} Simon, N. R. \& Schmidt, E. G. 1976, ApJ, 205, 162  
\bibitem []{} Stellingwerf, R. F. 1975, ApJ, 195, 441 
\bibitem []{} Stellingwerf, R. F. 1978, AJ, 83, 1184  
\bibitem []{} Stellingwerf, R. F. 1982, ApJ, 262, 330 
\bibitem []{} Stellingwerf, R. F., Gautschy, A. \& Dickens, R. J. 1987, ApJ, 
313, L75 (SGD)  
\bibitem []{} Stellingwerf, R. F. 1990, in The Numerical Modelling of Nonlinear
Stellar Pulsations, ed. J.R. Buchler (NATO ASI Ser. C302, Dordrecht: Kluwer), 27
\bibitem []{} Stothers, R. B. 1987, ApJ, 319, 260  
\bibitem []{} Udalski, A., Olech, A., Szymanski, M., Kaluzny, J., Kubiak, M., 
Mateo, M., \&  Krzemi\'nski, W. 1995, Acta Astron., 45, 433   
\bibitem []{} van Albada, T. S., \& Baker, N. 1973, ApJ, 185, 477 
\end{thebibliography}
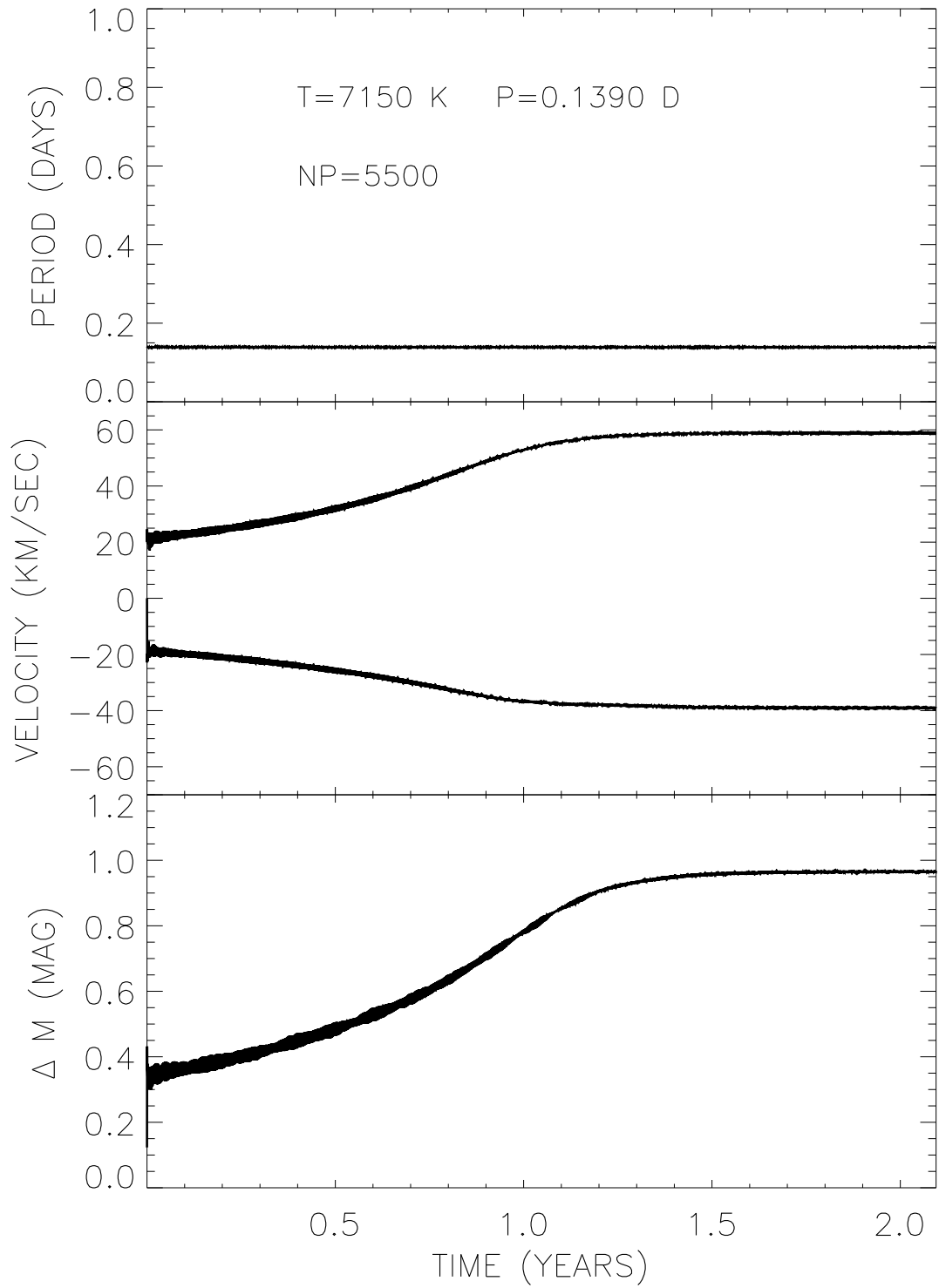
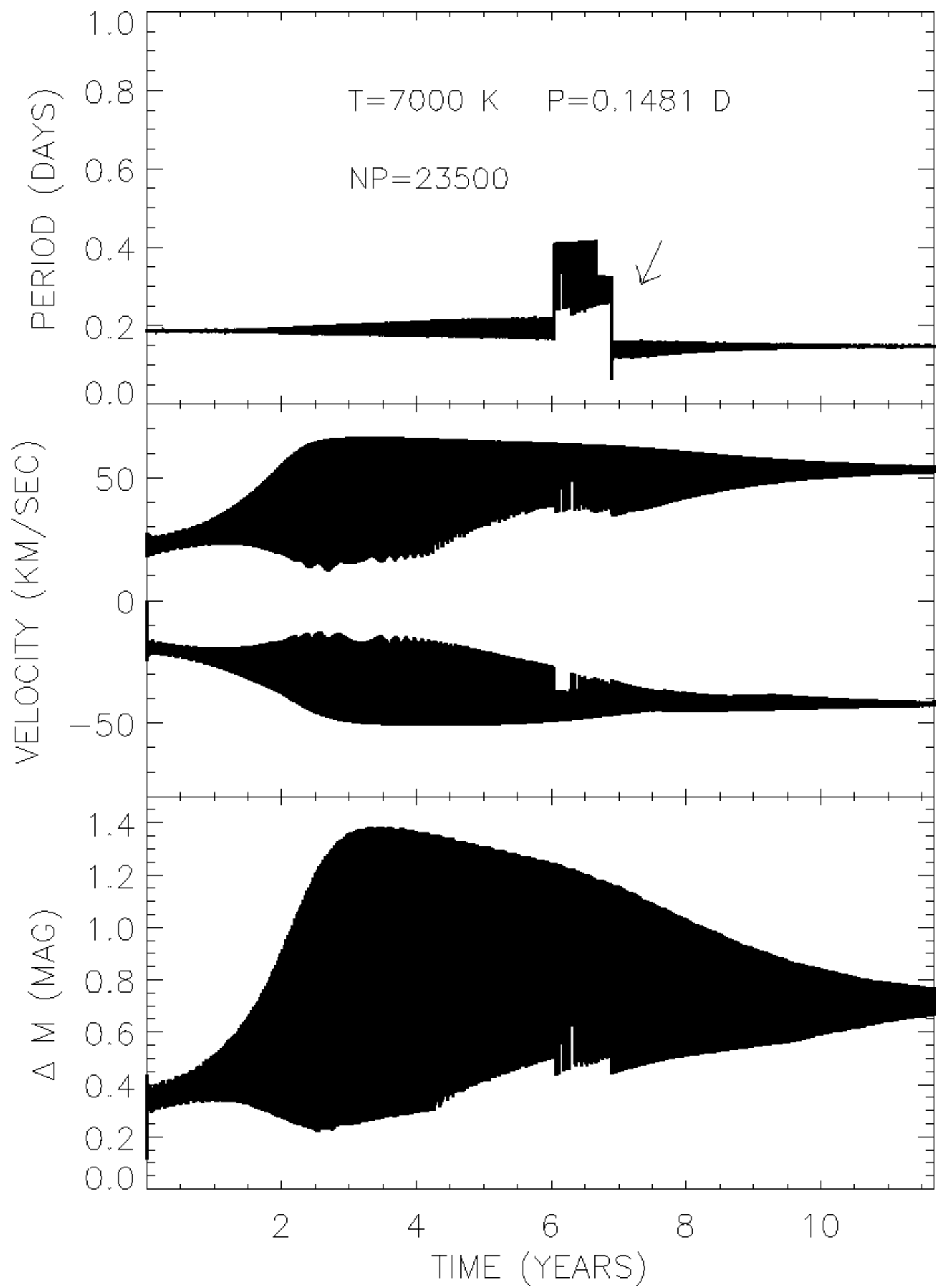
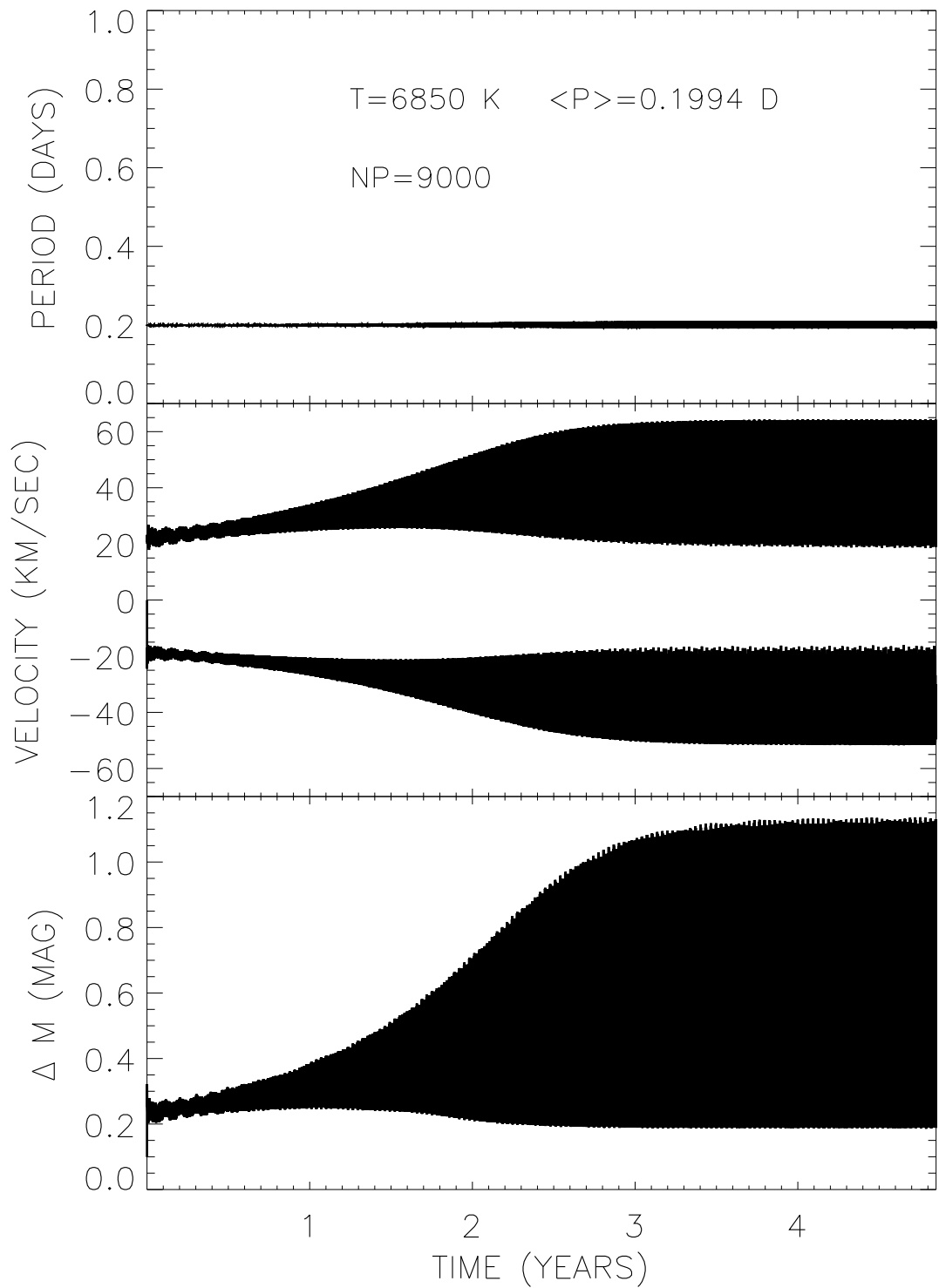
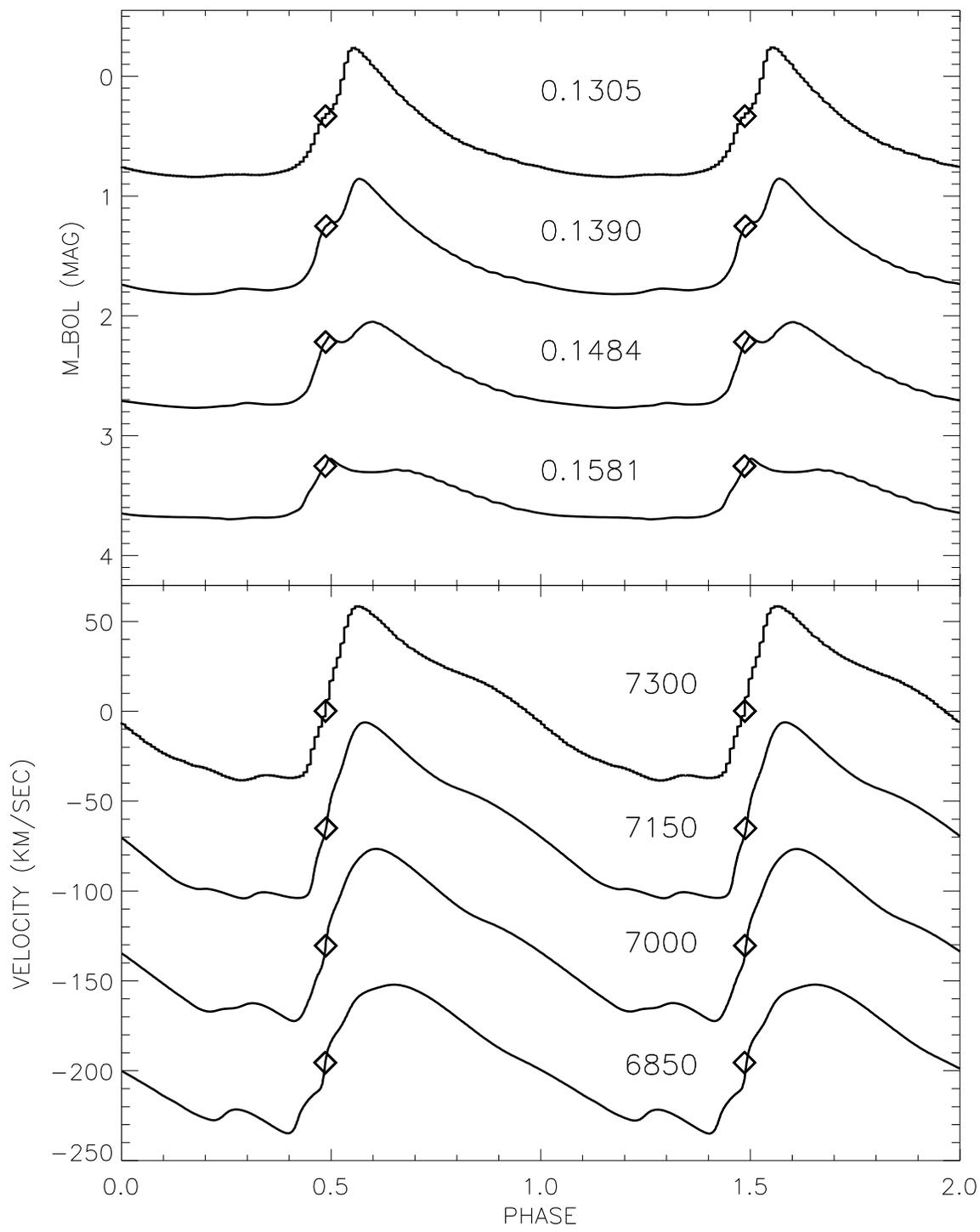
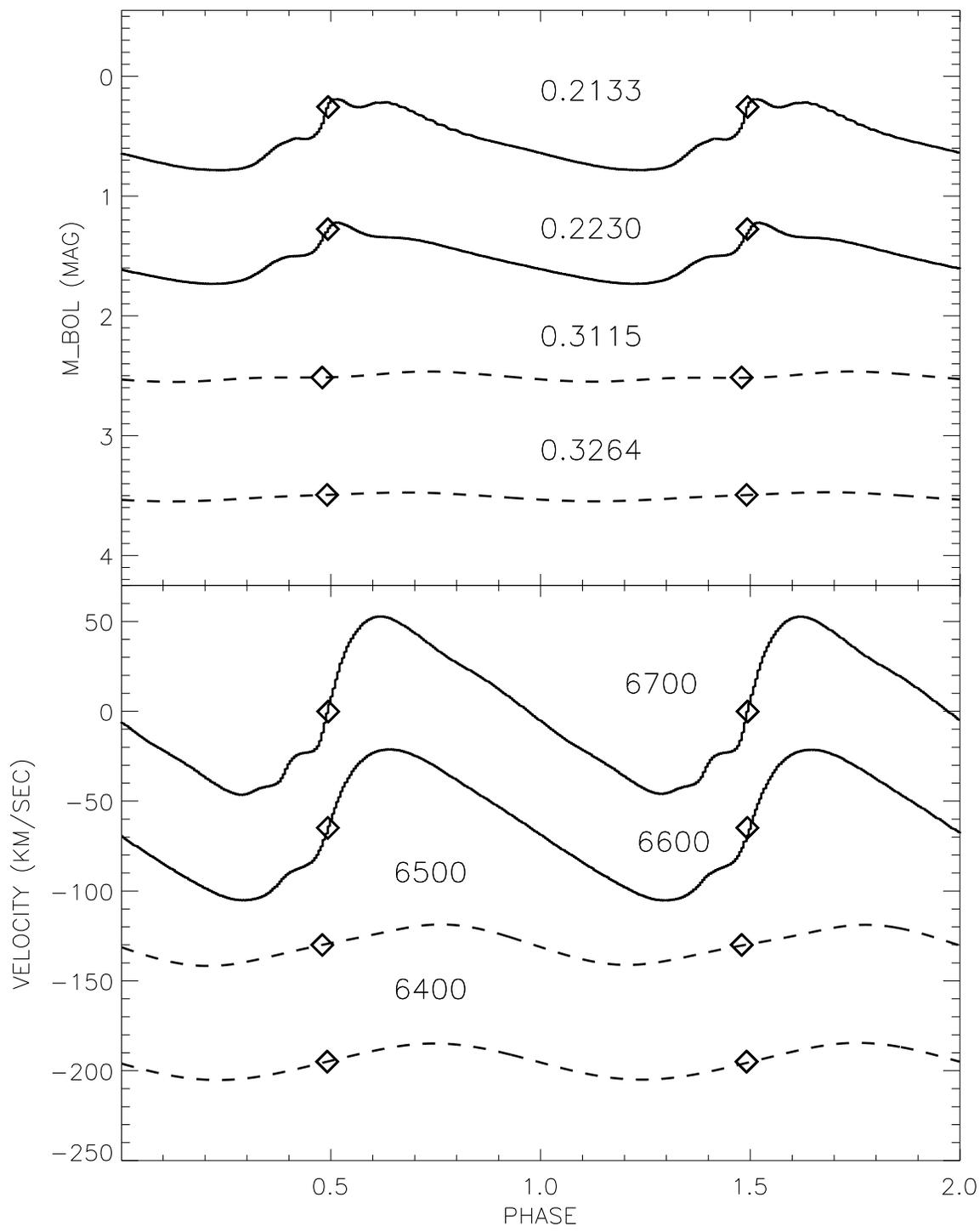
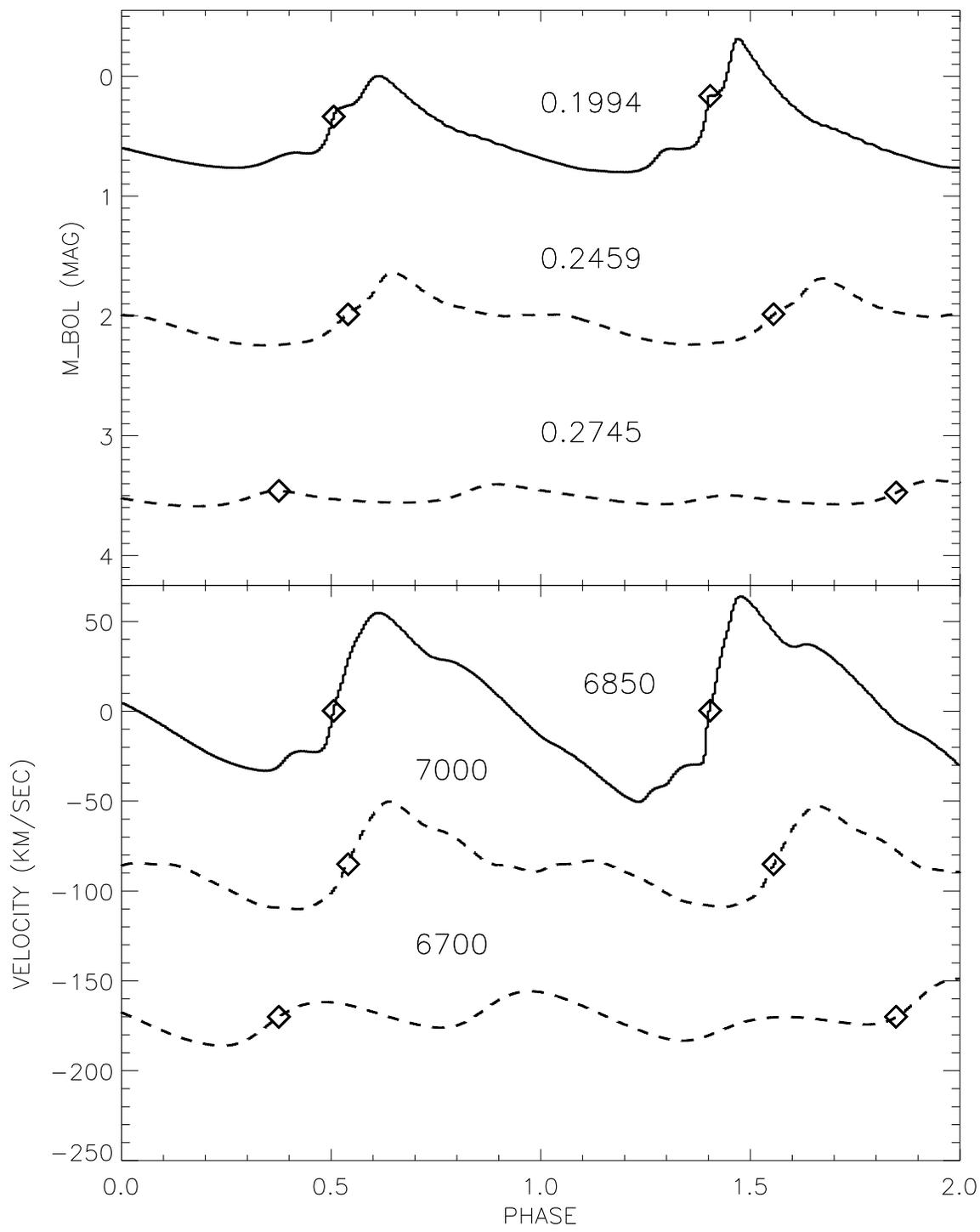
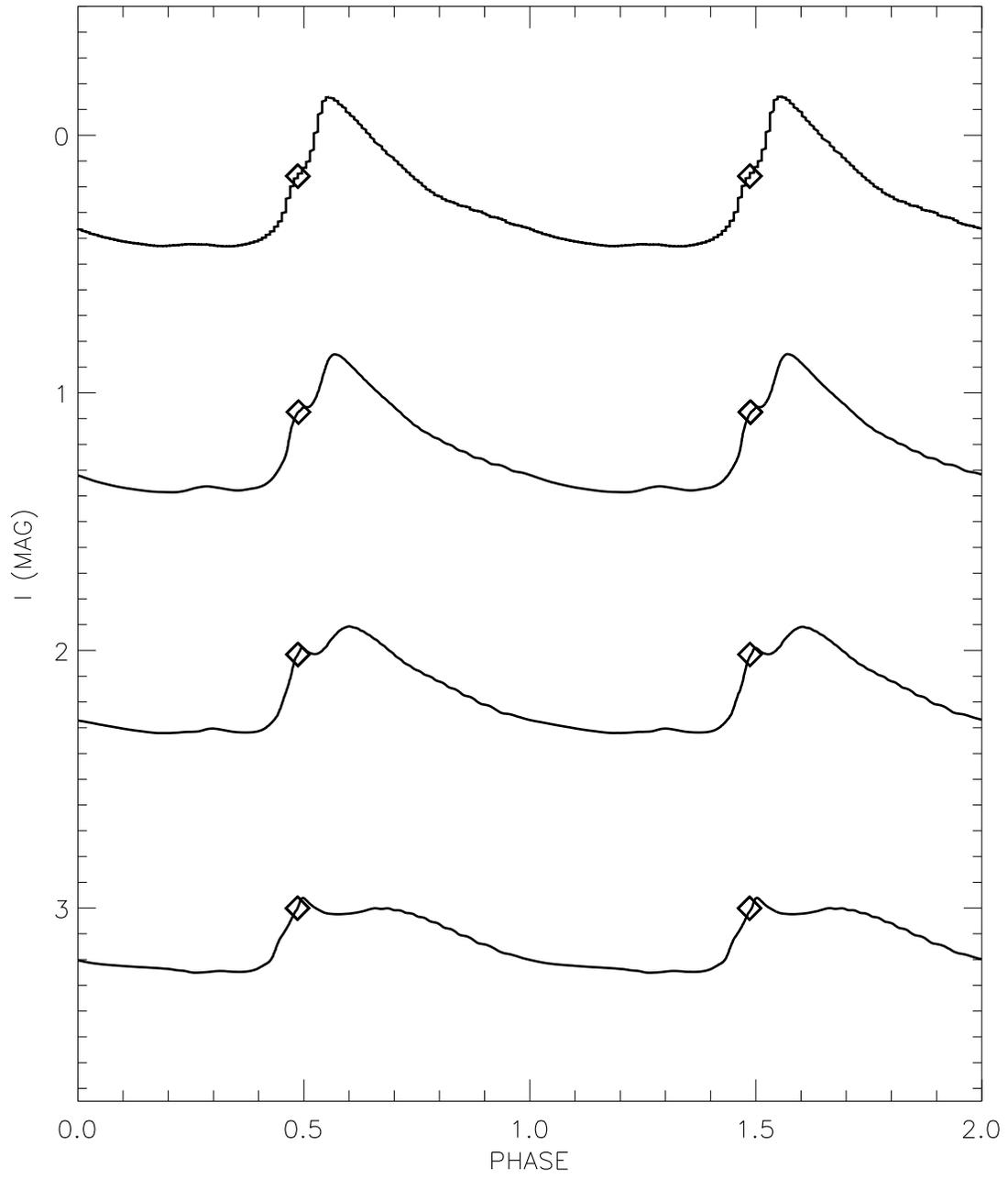
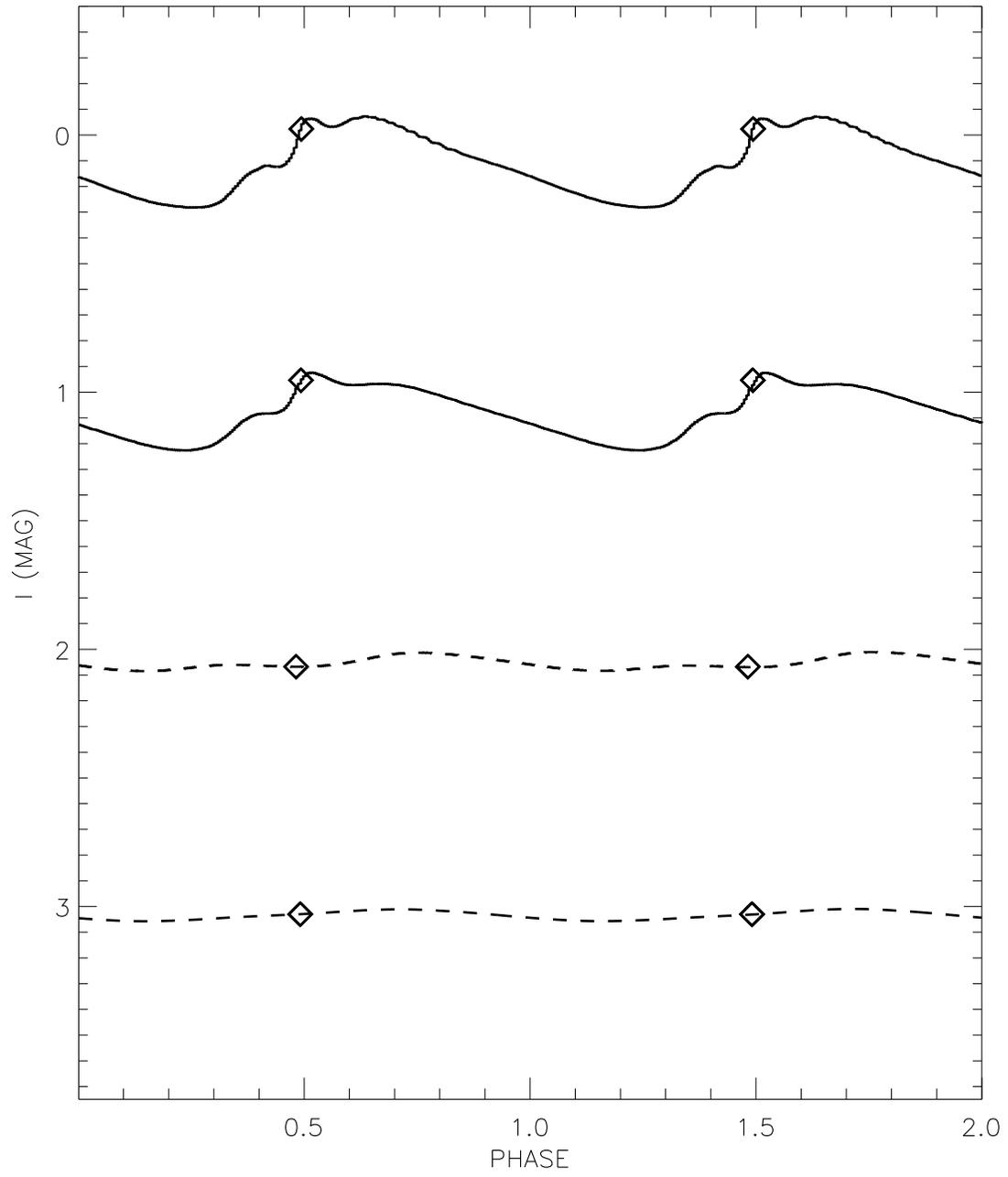
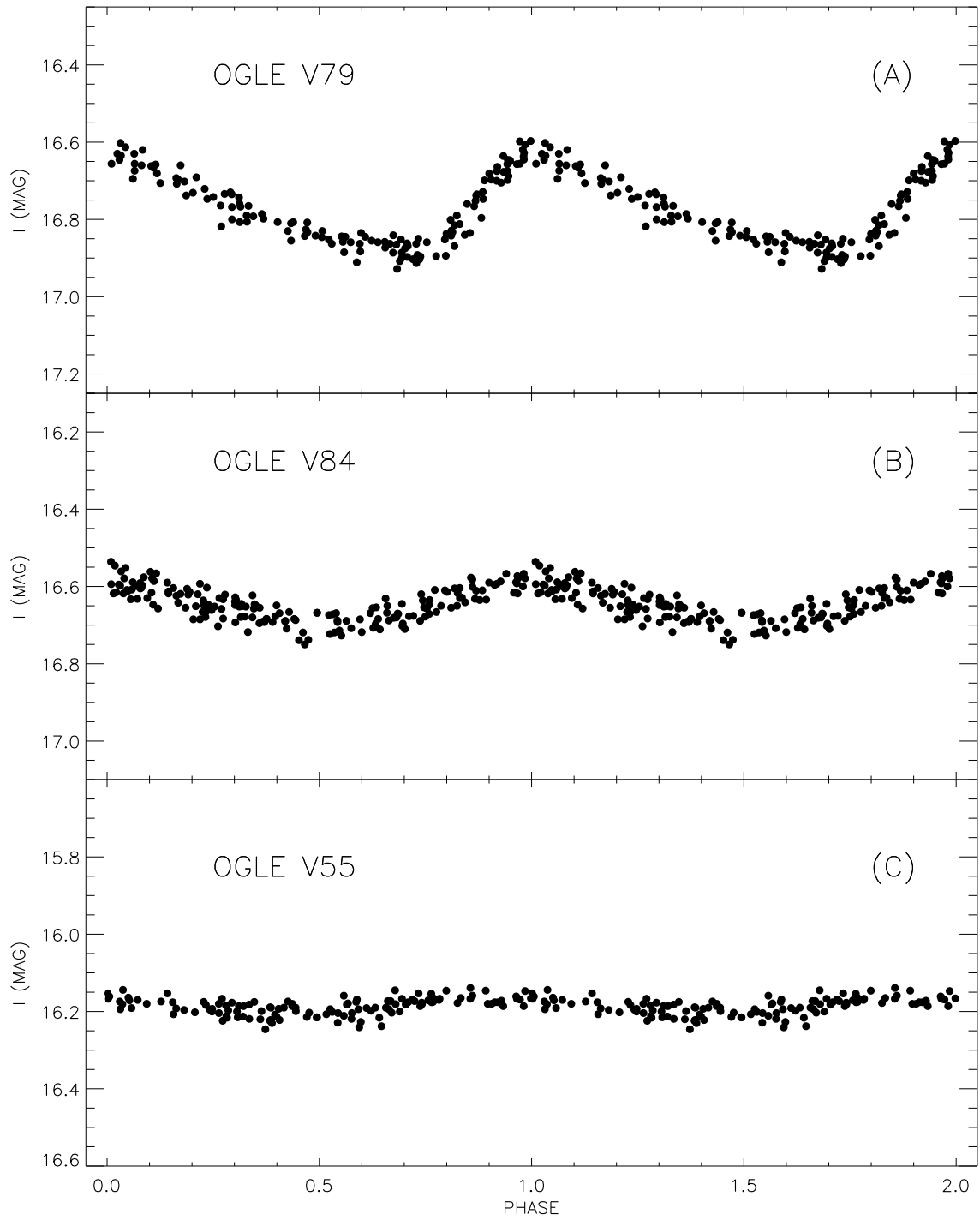
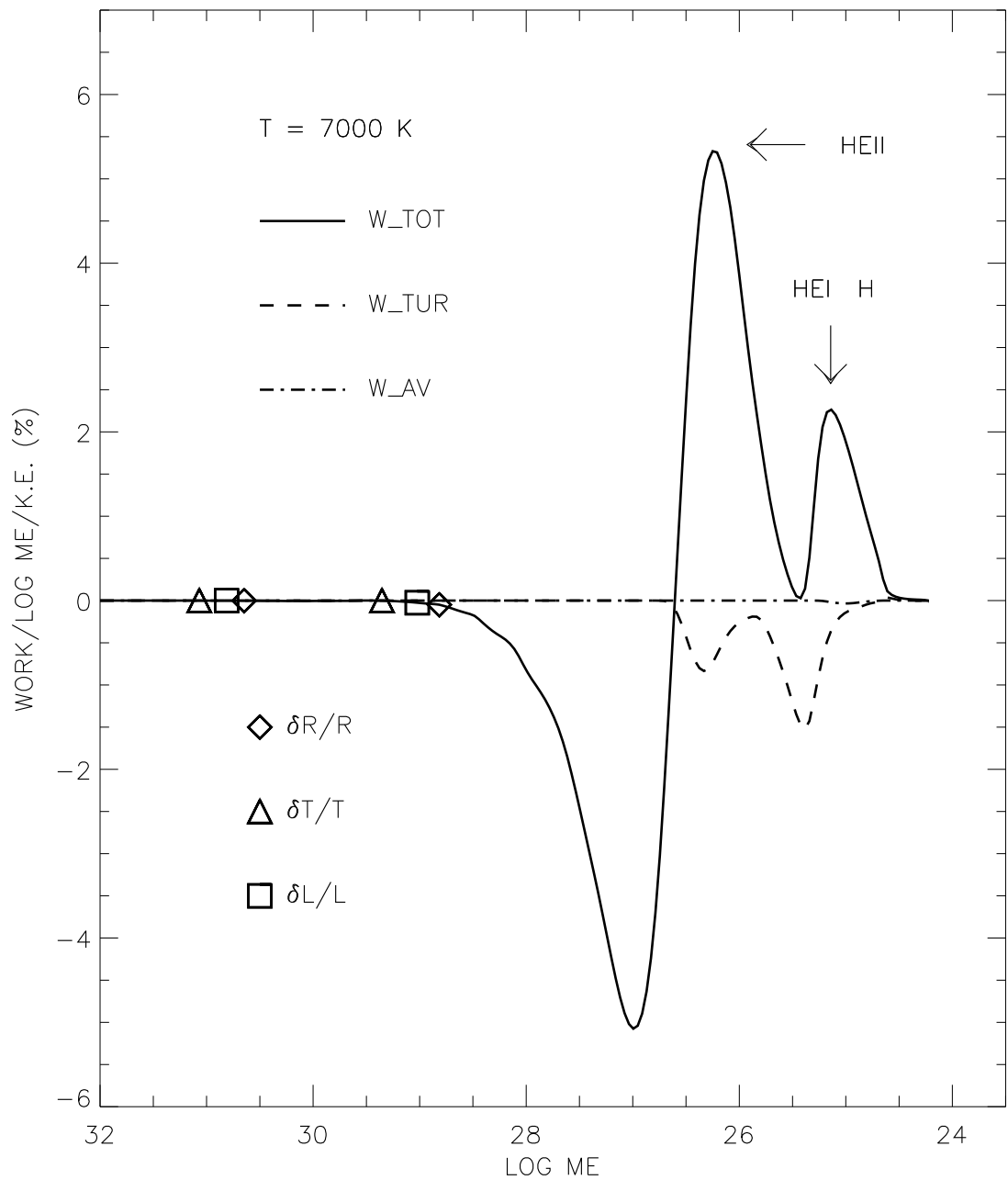
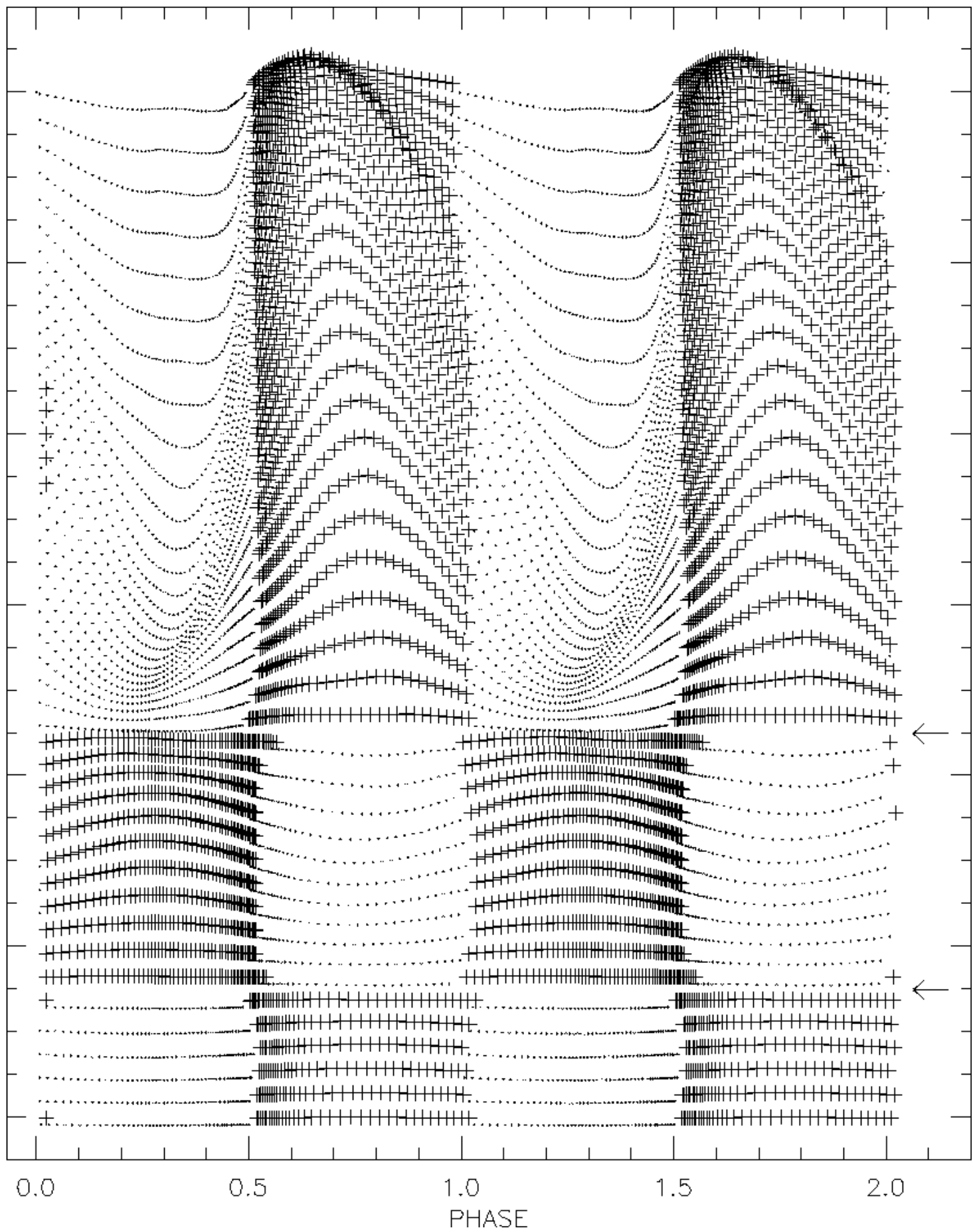
\end{document}